\documentclass[trackchanges]{aastex7}
\usepackage{CJK}

\begin{document}

\begin{CJK*}{UTF8}{gbsn}  

\title{Probing Proton versus Electron Heating and Energization during Magnetic Reconnection}

\author[orcid=0009-0009-7643-6103,sname='Yin']{Zhiyu Yin (尹志宇)}
\affiliation{Department of Physics, University of Maryland, College Park, MD 20742, USA}
\affiliation{IREAP, University of Maryland, College Park, MD 20742, USA}
\email[show]{zyin1@umd.edu}

\author[orcid=0000-0002-9150-1841,sname='Drake']{J. F. Drake}
\affiliation{Department of Physics, University of Maryland, College Park, MD 20742, USA}
\affiliation{IREAP, University of Maryland, College Park, MD 20742, USA}
\affiliation{Institute for Physical Science and Technology, University of Maryland, College Park, MD 20742, USA}
\affiliation{Joint Space Science Institute, University of Maryland, College Park, MD 20742, USA}
\email{drake@umd.edu} 

\author[orcid=0000-0002-5435-3544,sname='Swisdak']{M. Swisdak}
\affiliation{IREAP, University of Maryland, College Park, MD 20742, USA}
\affiliation{Joint Space Science Institute, University of Maryland, College Park, MD 20742, USA}
\email{swisdak@umd.edu} 

\begin{abstract}

The mechanisms controlling the relative heating and energization of electrons and protons during magnetic reconnection are explored. Simulations are carried out with the \textit{kglobal} model, which produces bulk heating and the extended powerlaw distributions of both species that have been documented in observations. The simulations have been carried out with a range of proton-to-electron mass ratios and upstream temperatures to isolate the factors that control energy gain. The simulations reveal that when the upstream temperatures of the two species are equal, the proton heating and energization exceeds that of electrons and that this is a consequence of the much larger energy gain of protons on their first entry into the reconnection exhaust. The effective energy gain of protons on exhaust entry scales as $m_iC_A^2$ since the protons counterstream at the Alfv\'en speed $C_A$ while the initial electron energy gain is smaller by the factor $(\beta_{e0}m_e/m_i)^{1/2}$. Since Fermi reflection during flux rope merger dominates energy gain in large-scale reconnecting systems and the rate of energy gain is proportional to energy, protons continue to gain energy faster than electrons for the duration of the simulations, leading to temperature increments of protons exceeding that of electrons and the non-thermal energy content of protons also exceeding that of electrons.  

\end{abstract}




\section{Introduction} 
\label{sec:intro}
Magnetic reconnection is responsible for the rapid conversion of magnetic energy in various environments, including solar flares \citep{lin71,Masuda94,Lin03,Benz17}, Earth's magnetosphere \citep{Dungey61,Sonnerup81}, and the solar wind \citep{Gosling05,Phan06,phan21}. During reconnection, magnetic energy is efficiently transferred to particles, producing both bulk heating that scales with $m_iC_A^2$ (the available magnetic energy per particle) \citep{Phan13a,Phan14,Oieroset23,Oieroset24,Oka25} and a significant number of non-thermal particles characterized by power-law tails in their distribution functions \citep{Lin03,Oieroset02,Krucker10,Gary18,Ergun20b,Desai25}. {\it In situ} measurements of bulk heating have revealed that protons gain significantly more energy than electrons and there is some evidence that protons may also dominate energy gain in the non-thermal component during magnetotail reconnection \citep{Ergun20,Rajhans25}. In the case of solar flares, measurements of energetic protons do not extend below around an MeV so the total energy content of protons is uncertain \citep{Lin03,Emslie12}. However, recent observations of line broadening in flares suggests ion heating in flares might also exceed that of electrons \citep{Russell25}. 

Magnetic reconnection produces bent magnetic field lines that expand outward due to their tension force to form an Alfv\'enic exhaust \citep{Parker57,Petschek64,Sato79}. The exhaust carries energy away from the x-line and transfers energy from the magnetic field to the surrounding plasma \citep{Lin93}. Reconnecting current sheets, however, tend to breakup and form multiple x-lines and associated flux ropes \citep{Biskamp86,Drake06,Loureiro07,Bhattacharjee09,Daughton11} and observations support the multi-x-line picture of reconnection \citep{Chen08,Phan24} The dominant mechanism driving particle energy gain during magnetic reconnection is Fermi reflection in growing and merging magnetic flux ropes \citep{Drake06,Oka10,Dahlin14,Guo14,Li19,Zhang21}. The mechanism produces power-law tails in both electron \citep{Arnold21} and ion \citep{Zhang21,Yin24b} distributions. Particles gain energy through repeated reflections in contracting magnetic field lines, thereby leading to the observed power-law tails in particle energy distributions \citep{Drake06,Drake13,Li19,Zhang21}. 
The rate of energy gain from Fermi reflection is proportional to a particle's energy \citep{Drake06,Drake13} and is therefore greatest for the most energetic particles. For this reason, the most energetic particles gain the most energy, which facilitates the formation of extended powerlaw tails. The \textit{kglobal} model \citep{Arnold19, Drake19, Yin24}, which is designed to describe reconnection in macroscale systems by ordering out all kinetic scales, was the first fully self-consistent model to produce the extended powerlaw tails documented in observations \citep{Arnold19, Drake19}. 

The \textit{kglobal} model was recently upgraded to include particle protons \citep{Yin24} and simulations using this model revealed that the energy content of energetic protons exceeded that of electrons \citep{Yin24b} even when both species start with equal initial temperatures. Earlier particle-in-cell (PIC) simulations also explored the relative heating of electrons and protons during reconnection \citep{Haggerty15}. These simulations established that the proton temperature gain exceeds that of the electrons during magnetic reconnection. However, the underlying mechanisms that facilitate proton versus electron heating and energization have not been identified and the PIC simulations were unable to match the observed non-thermal spectra of the two species. 

In this paper, we present simulation results that capture the heating and energization of protons and electrons during magnetic reconnection and that address the physics basis for the differences in the heating and energization between the two. Section \ref{sec:modelsetup} details the simulation setup, Section \ref{sec:results} presents the key findings, and Section \ref{sec:conclusion} summarizes our conclusions and discusses their implications for understanding reconnection-driven particle energization and its potential applications.

\section{Simulation Model Setup}
\label{sec:modelsetup}
The simulations include four distinct plasma species: fluid protons and fluid electrons (which collectively form the MHD-like backbone), and particle protons and electrons which are represented by macro-particles that move through the fluid grid.  The particles are treated in the guiding-center limit, thus eliminating the need to resolve their respective Larmor radii. The simulations were performed within a two-dimensional (2D) spatial domain while motions are allowed in three  directions. Particles move across the magnetic field with their $\mathbf{E}\times\mathbf{B}$ drift and along the local magnetic field at their parallel velocity. The reconnecting component of the upstream magnetic field \( B_0 \) (along the $x$ direction) and the total proton density (the combined number density of particle and fluid protons) \( n_{i0} \) serve as normalization parameters by defining the Alfv\'en speed \( C_{A0} = B_0 / \sqrt{4\pi m_i n_{i0}} \). Because kinetic scales are excluded in the model, lengths are normalized to an arbitrary macroscale \( L_0 \) and time scales are normalized to \( \tau_A = L_0 / C_{A0} \).  Both temperatures and particle energies are normalized to \( m_i C_{A0}^2 \), which means that the intrinsic heating of the plasma scales with this parameter, as revealed by {\it in situ} observations. The perpendicular electric field in the simulations follows the MHD scaling \( C_{A0} B_0 / c \), while the parallel field scales as \( m_i C_{A0}^2 / (e L_0) \). Although the parallel electric field is small compared to the perpendicular component, the energy associated with the parallel potential drop over the scale \( L_0 \) is of the order of \( m_i C_{A0}^2 \), making it comparable to the available magnetic energy per particle.

The proton-to-electron mass ratio is varied from $25$ to $400$ to investigate the parameters that control the relative energy gain of the electrons and protons. The simulations are initialized with constant densities and pressures in a force-free current sheet with periodic boundary conditions.  It has previously been shown that particle heating and energization is insensitive to the size of the simulation domain and therefore the periodic boundary conditions \citep{Arnold21,Yin24b}. Thus, $B = B_0 \tanh(y/w) \hat{x} + (B_0^2\,\text{sech}^2(y/w) + B_g^2)^{1/2} \, \hat{z}$  where $B_g$ is the asymptotic out-of-plane magnetic field (the guide field) and $w$ is the width of the current sheet, which is set to 0.005$L_0$. The initial total density of electrons ($n_e$) and protons ($n_i$) is normalized to unity, with  particles comprising $25\%$ of the density ($n_{ep}$ for particle electrons and $n_{ip}$ for particle protons) and the remaining $75\%$ in the fluid component ($n_{ef}$ for fluid electrons and $n_{if}$ for fluid protons). In earlier simulations we showed that results of the simulations are insensitive to this fraction \citep{Yin24b}. The simulations are conducted on grids with a resolution of $2048 \times 1024$ grid points as shown in Table \ref{tab:initial_setting}. Each grid cell initially holds 100 particles per species. 

\begin{deluxetable*}{ccccc}
\tablewidth{0pt}
\tablecaption{Parameters for Simulation Domains\label{tab:initial_setting}}
\tablehead{
  \colhead{Grid Points} & 
  \colhead{$2048 \times 1024$}
}
\startdata
Time Step (in $\tau_A$) & $1 \times 10^{-4}$ \\
Proton Number Density Diffusion ($D_n$) & $5.25 \times 10^{-5}$ \\
Proton Pressure Diffusion ($D_p$) &  $5.25 \times 10^{-4}$ \\
Hyperviscosity ($\nu_B$, $\nu_{nv}$, $\nu_n$, and $\nu_p$) & $1.05 \times 10^{-8}$ \\
Effective Lundquist Number ($S_\nu$) & $9.5 \times 10^{7}$ \\
\enddata
\end{deluxetable*}

In our simulations, diffusion and hyperviscosity terms are included to ensure numerical stability while reducing any high-frequency noise arising at the grid scale.  The diffusion coefficients, denoted as \(D_n\) for number density diffusion and \(D_p\) for pressure diffusion, are set to the values listed in Table \ref{tab:initial_setting}. 
A hyperviscosity  \(\nu\) rather than a resistivity is included in the magnetic field evolution equation to facilitate reconnection while minimizing dissipation at large scales. It is applied as a fourth-order Laplacian term (\(\nabla^4\)) in the equation governing the evolution of the magnetic field. The same viscosity is included in the evolution equations for the fluid proton flux, fluid proton number density and fluid proton pressure. The effective Lundquist number \(S_\nu = C_A L_0^3 / \nu\) associated with the hyperviscosity is varied to change the effective system size (the ratio of the macro to the dissipation scale).
The hyperviscosity coefficients, denoted as \(\nu_B\), \(\nu_{nv}\), \(\nu_n\), and \(\nu_p\), as well as $S_\nu$, are set to the values given in Table \ref{tab:initial_setting}. 

To explore electron and proton heating and energization, we perform reconnection simulations and evaluate the energy spectra and temperature increments of both species. We limit the analysis to the particle electrons and protons. In showing the energy spectra, we present data from the entire computational domain. This analysis increases the statistics for the highest energy particles. The evaluation of the average temperature increments for comparison with observations is non-trivial because reconnection involves the formation and merger of multiple flux ropes, which have complex spatial structure. Our procedure is to evaluate the highest temperature increment on the grid at a particular time and average the temperature over locations with temperatures above 0.75 of this peak value that are also within the separatrix that extends furthest from the center of the current sheet. This avoids including upstream plasma within the separatrix that has not undergone heating in the average. 

We begin the simulations with a reference case where both species have equal upstream temperatures: $T_e = T_i = 0.0625\, m_i C_A^2$. In subsequent simulations we explore the impact of the relative temperatures of the upstream particles by keeping the proton temperature fixed at $T_i = 0.0625\, m_i C_A^2$ while varying the upstream electron temperature across a range of values: $T_e = 5 \times 0.0625\, m_i C_A^2$, $2 \times 0.0625\, m_i C_A^2$, and $0.5 \times 0.0625\, m_i C_A^2$. This setup enables a direct comparison of energy gain under different initial electron-to-proton upstream temperature ratios.
\section{Simulation Results} \label{sec:results}

\begin{figure*}[ht!]
\centering
\includegraphics[width=\columnwidth]{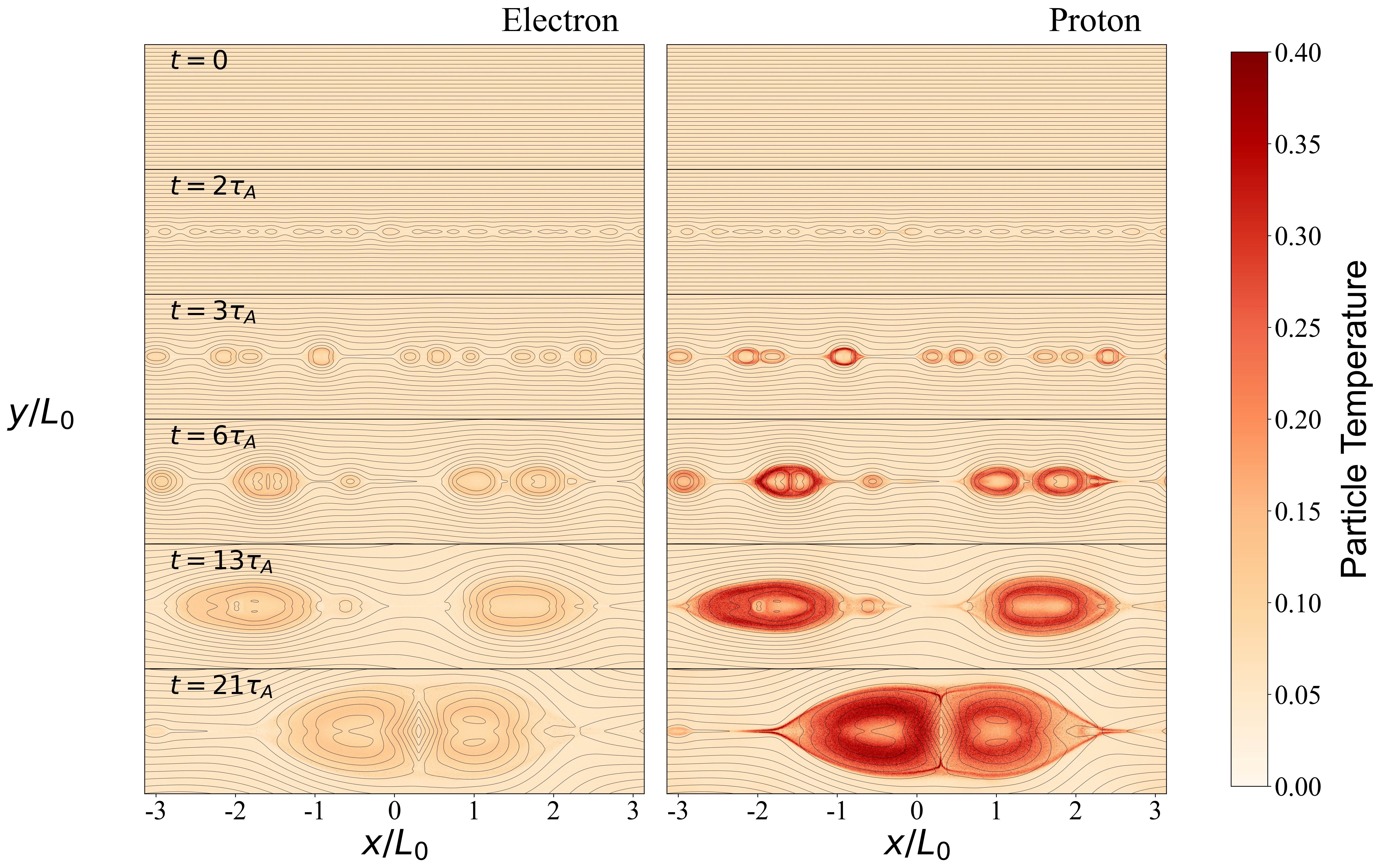}
\caption{The temporal evolution of particle electron (lsft) and proton (right) temperatures in the \( x\text{--}y \) plane for \( B_g/B_0 = 0.25 \). Six time steps are shown: \( t/\tau_A = 0, 2, 3, 7, 13, \) and \( 21 \). In each panel, magnetic field lines are overlaid in black.
\label{fig:evo}}
\end{figure*}

In our simulations, reconnection is facilitated by hyperviscosity, which results in the generation of small-scale flux ropes at early time. These structures gradually merge, eventually producing large, system-scale flux ropes. In Figure \ref{fig:evo} are the results of simulations with \( B_g/B_0 = 0.25 \), displaying both electron and proton particle temperatures in the \( x\text{--}y \) plane at six different times: \( t/\tau_A = 0, 2, 3, 6, 13, \) and \( 21 \). The color bars in these panels are normalized to $m_iC_{A0}^2$ as are the temperatures presented in subsequent figures.

At \( t = 0 \), the magnetic field reverses across a uniform current sheet, and both particle electrons and protons have the same temperature, as shown in the top panel of Figure \ref{fig:evo}. As the system evolves, magnetic reconnection begins at multiple locations, and the reconnected magnetic field lines are convected away from the X-points. Starting from \( t = 2\tau_A \), the proton temperature becomes noticeably higher than that of the electrons. The contraction of magnetic islands along the current sheet leads to the energization of electrons and protons trapped within these structures. As the simulation progresses, the islands grow and merge, eventually forming a single, large magnetic island at late time. As shown in Figure \ref{fig:evo}, the proton temperature is significantly higher than the electron temperature once reconnection commences.

\begin{figure*}[ht!]
\centering
\includegraphics[width=\columnwidth]{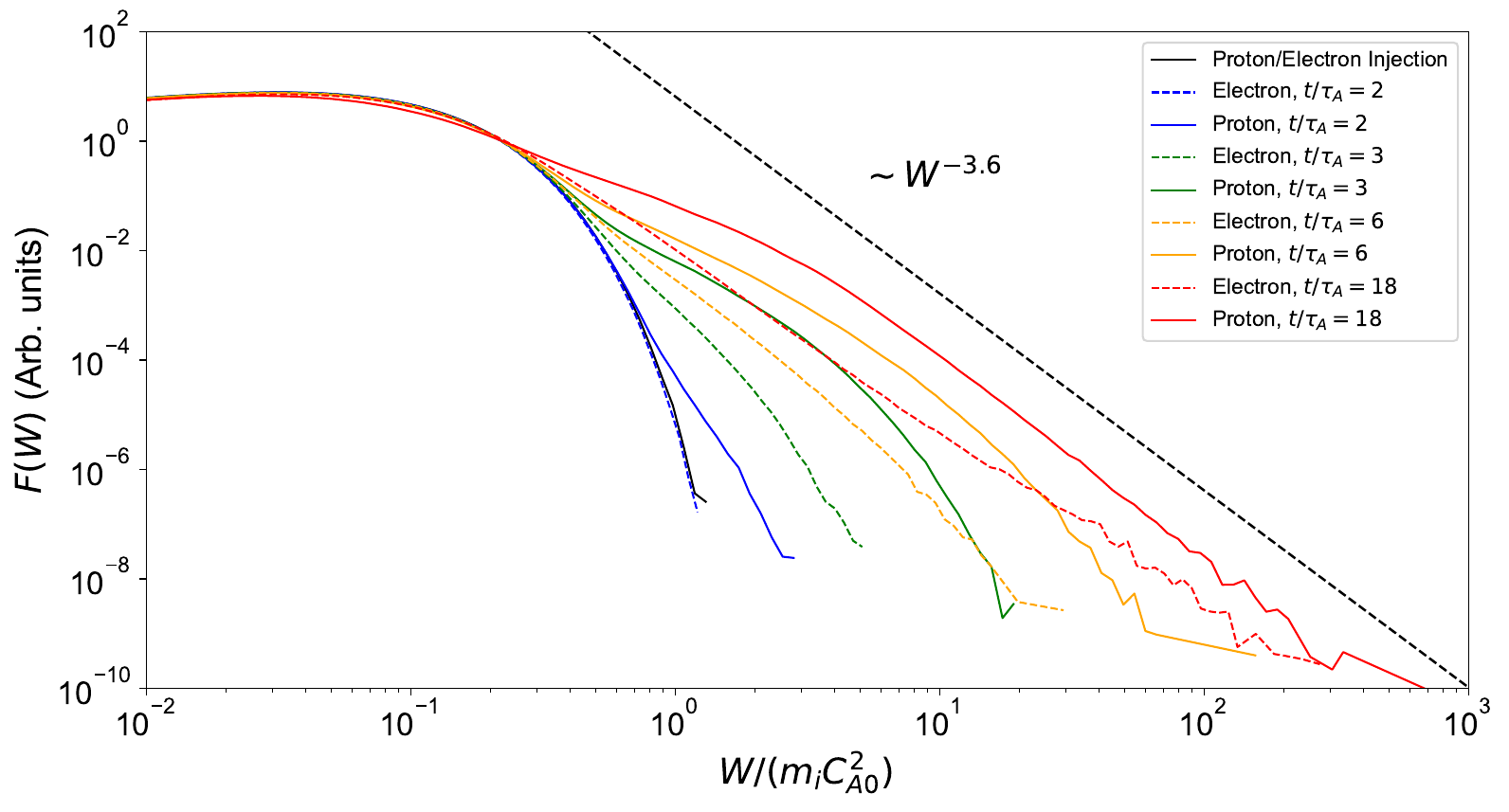}
\caption{Time evolution of the energy spectra of electrons (dashed lines) and protons (solid lines) during the simulation with equal initial temperatures. The initial spectrum （proton/electron）is shown as a black solid line. Subsequent spectra are plotted at \( t/\tau_A = 2 \), 3, 6, and 18. A black dashed line line with a slope of -3.6 matches the slope of the late-time proton spectrum. The late time powerlaw index of the electrons is around -3.2.
\label{fig:pe_spec_evo}}
\end{figure*}

The particle spectra from the same simulation are displayed in  Fig.~\ref{fig:pe_spec_evo} at several times during the simulation (electrons in dashed lines and protons in solid lines). The data is taken from the entire simulation domain. As early as \(t = 2\tau_A\), protons already exhibit greater energization than electrons. They maintain their energy advantage over electrons during the subsequent evolution.

\begin{figure*}[ht!]
\centering
\includegraphics[width=6.0in]{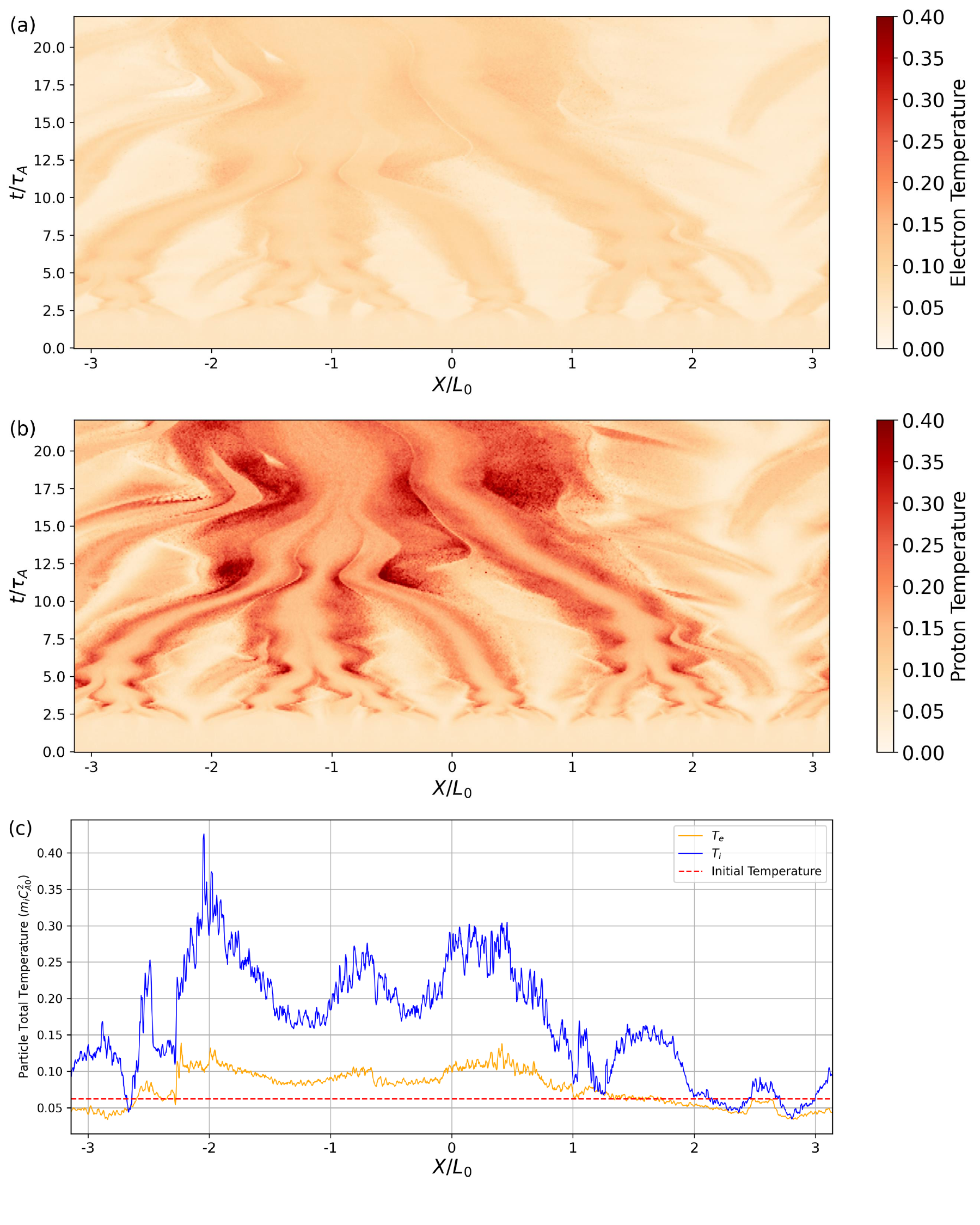}
\caption{Temperature evolution and late-time profile for a simulation with $T_{e(\mathrm{initial})} = T_{i(\mathrm{initial})}$ and $m_i/m_e = 25$. Panels (a) and (b) show the space–time diagrams of the parallel electron and proton temperatures, respectively, along the center of the current sheet. The horizontal axis represents position $X/L_0$, and the vertical axis shows time normalized to the Alfvén time $\tau_A$. The color scale indicates normalized temperature, with redder regions corresponding to higher temperatures. Panel (c) presents a late-time 1D cut through the center of the current sheet, showing parallel electron temperature $T_e$ (orange) and proton temperature $T_i$ (blue). The red dashed line marks the initial temperature for both species, $T_0 = 0.0625\,m_i C_{A0}^2$. 
\label{fig:combine_cut}}
\end{figure*}

This persistent energy gap between protons and electrons is further illustrated in Figure~\ref{fig:combine_cut}. Panels (a) and (b) show the space–time evolution of the parallel electron and proton temperatures in a cut along the center of the current sheet. The flux ropes reveal themselves as adjacent regions of higher temperature. Flux ropes merge as reconnection develops. While both species are heated as reconnection progresses, proton heating is stronger and becomes evident as early as $t \approx 2\,\tau_A$. Shown in Panel (c) are cuts at late time along the center of the current sheet of the bulk temperature profiles of electrons and protons. The proton temperature significantly exceeds the electron temperature across the domain, typically reaching a sizable fraction of $m_i C_{A0}^2$, while the electrons exhibit only modest heating. 

\begin{figure*}[ht!]
\centering
\includegraphics[width=6.0in]{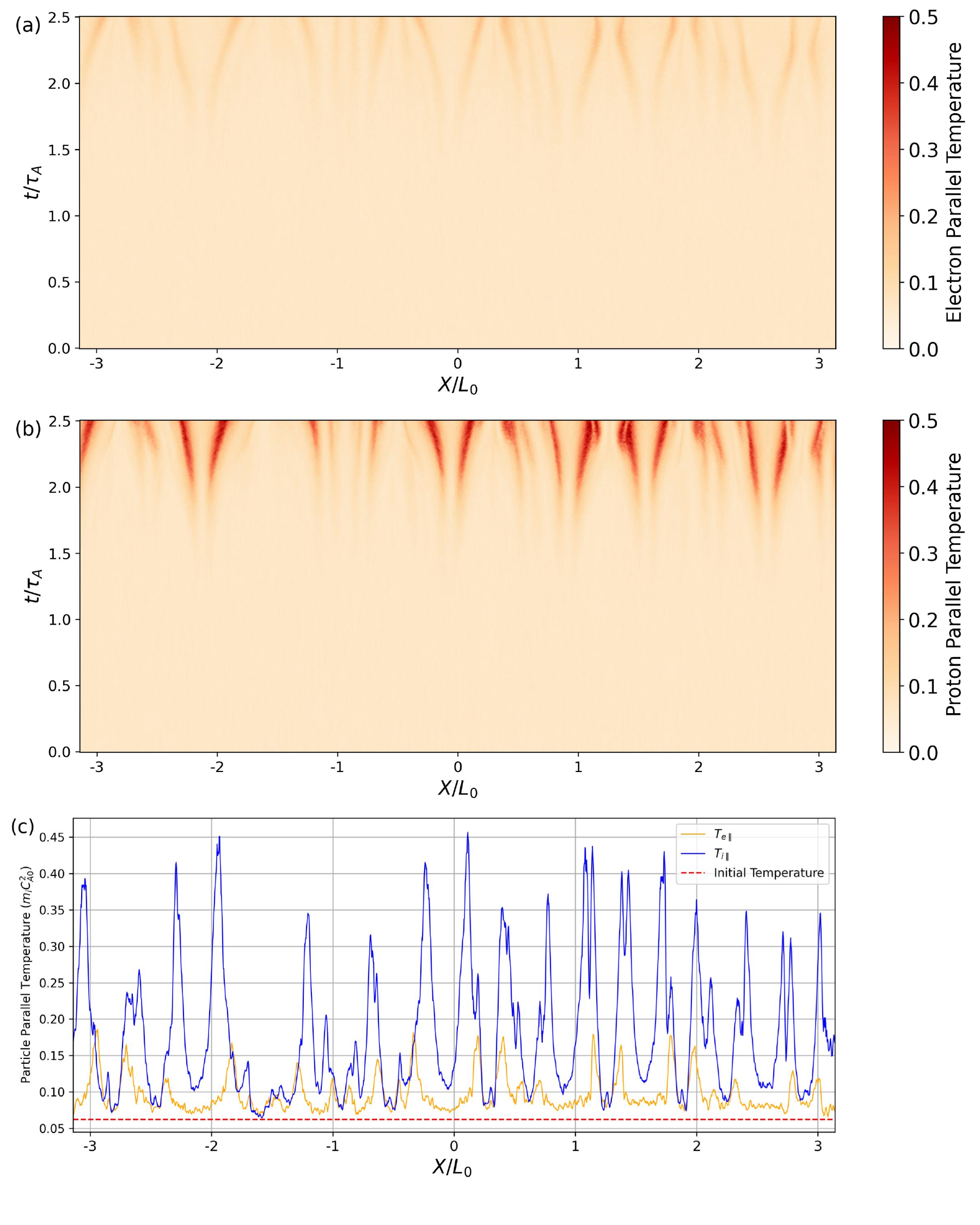}
\caption{Early phase evolution of particle temperatures during magnetic reconnection for the simulation in Figure \ref{fig:evo}. In (a) and (b) the space–time evolution of the parallel electron and proton temperatures, respectively, along the center of the current sheet.  Panel (c) shows cuts of the parallel temperature of protons (blue) and electrons (orange) along the center of the current sheet at $t = 2.5\,\tau_A$. The red dashed line marks the initial temperature.
\label{fig:early_time}}
\end{figure*}

It is evident from Figures \ref{fig:evo}-\ref{fig:combine_cut} that protons gain much more energy at early time and maintain that advantage throughout the reconnection simulation. To better understand these early-stage dynamics, we show the very early space–time evolution of electron and proton parallel temperatures along the center of the current sheet in panels (a) and (b) of Fig.~\ref{fig:early_time}, respectively. The diverging regions of rising temperature correspond to plasma being heated and flowing away from multiple x-lines.  Notably, protons exhibit more pronounced heating than electrons during this early phase, indicating that the protons are gaining more energy than electrons as they are injected into the outflow exhausts from reconnecting x-lines. 

The preferential energy gain of protons over electrons during magnetic reconnection can be quantitatively explained by the Fermi reflection mechanism. In this process, particles reflect off moving magnetic fields in reconnection outflows or during flux rope contraction. They gain a velocity increment of approximately $\Delta v \sim 2 C_A$ along the magnetic field direction. Although both electrons and protons receive a similar velocity kick, the resulting energy gain is mass-dependent. For a single reflection, the energy change is given by

\begin{equation}
\Delta \mathcal{E} = \frac{1}{2} m \left[ \left(v_{\parallel,0} + 2 C_A\right)^2 - v_{\parallel,0}^2 \right] 
= 2 m C_A v_{\parallel,0} + 2 m C_A^2
\label{eqn:fermi_proton}
\end{equation}
where $m$ is the particle mass and $v_{\parallel,0}$ is the initial parallel velocity. For protons with sub-Alfv\'enic upstream thermal velocities, the second term dominates and the energy gain during a single encounter with a reconnection exhaust scales like $m_iC_A^2$. For electrons, where thermal velocities upstream can exceed the Alfv\'en speed the first term dominates and scales as
\begin{equation}
    \Delta \mathcal{E}\sim \left(\beta_{e0}\frac{m_e}{m_i}\right)^{1/2}m_iC_A^2\ll m_iC_A^2.
    \label{eqn:fermi_electron}
\end{equation}

Thus, the much larger energy gain of protons compared to electrons at early time shown in Figure \ref{fig:early_time} is consistent with expectations based on Fermi reflection as particles first enter into reconnection exhausts.   Panel (c) in Fig. \ref{fig:early_time} shows a 1D cut of the parallel proton and electron temperatures along the center of the current sheet at $t = 2.5\,\tau_A$. The parallel proton temperature $T_i$ (blue) exhibits strong peaks centered within reconnection exhausts and reaching values above $0.4\,m_i C_{A0}^2$, indicating localized and efficient proton energization and consistent with Eq.~(\ref{eqn:fermi_proton}). In contrast, the electron parallel temperature $T_e$ (orange) shows only a moderate increase, with increments above their initial temperature (red dashed line) of around $0.1\,m_i C_{A0}^2$.  Thus, the electron energy gain is smaller by around a factor of 5 compared with protons, consistent with the scaling in Eq.~(\ref{eqn:fermi_electron}) for $m_i/m_e=25$.

It is important to note that the temperatures in Figure \ref{fig:early_time} are parallel temperatures so that their values can be compared with the expectations from Fermi reflection, which drives parallel heating. The perpendicular temperatures are much smaller and are actually reduced compared with their initial values because of the conservation of the magnetic moment $\mu=mv_\perp^2/2B$ and the reduction of $B$ in the reconnection exhaust. Thus, the temperatures at $t=2.5\,\tau_A$ in Figure \ref{fig:combine_cut} are smaller than those in Figure \ref{fig:early_time} because $T_\perp\ll T_\parallel$ within the exhausts (recall that the total temperature is defined as the average of the three components: $T_{\mathrm{tot}} = (T_\parallel + 2T_\perp)/3$).

\begin{figure*}[ht!]
\centering
\includegraphics[width=\columnwidth]{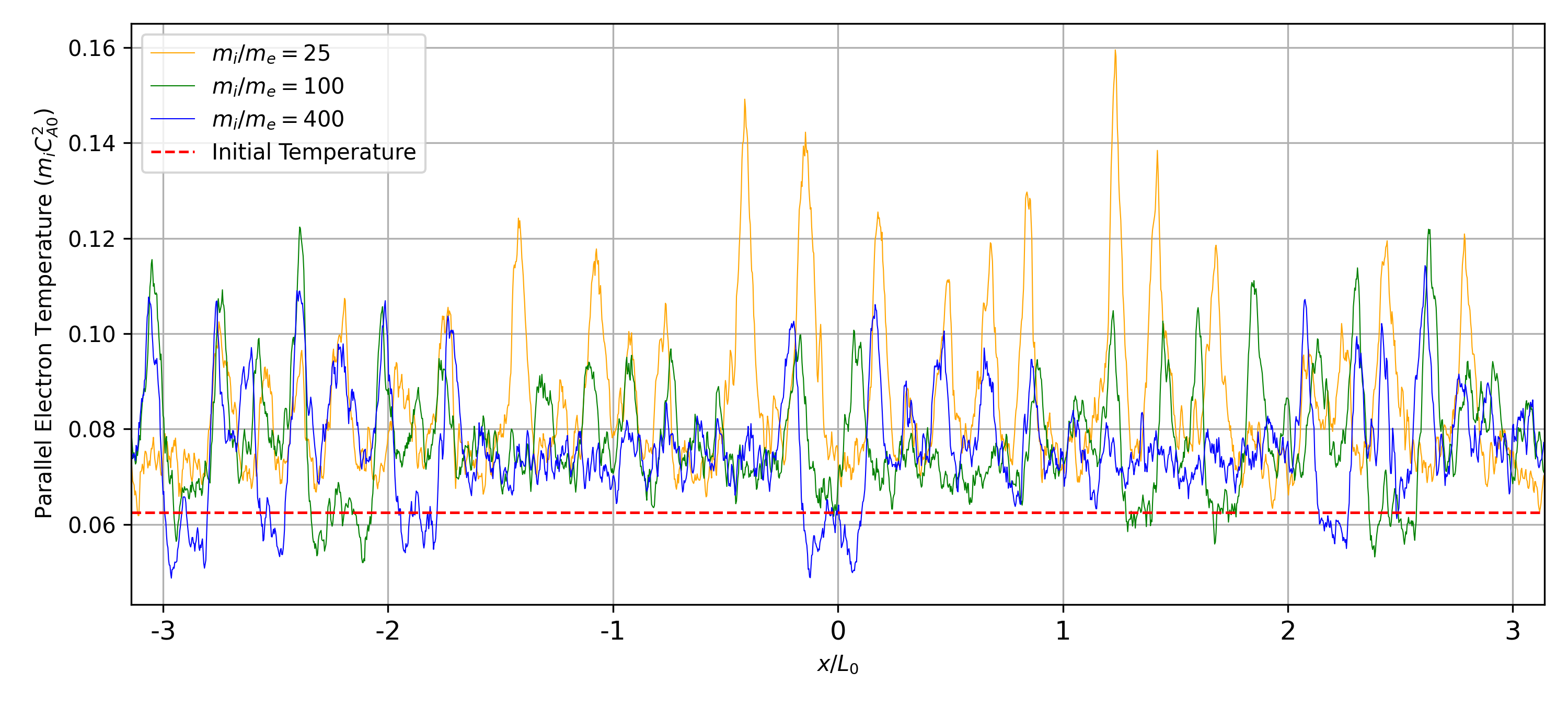}
\caption{Parallel electron temperature profiles at the current sheet at $t = 2.2\,\tau_A$ for three different mass ratios: $m_i/m_e = 25$ (orange), $m_i/m_e = 100$ (green), and $m_i/m_e = 400$ (blue), with the ion mass $m_i$ held constant in all simulations. The red dashed line denotes the initial electron temperature, $T_{e0} = 0.0625\, m_i C_{A0}^2$. The horizontal axis is the spatial coordinate $x$, normalized to the characteristic length scale $L_0$. Results show that heavier electrons (i.e., lower mass ratios) gain more energy than lower mass electrons during their initial injection into reconnection exhausts, consistent with Fermi reflection scaling where energy gain is proportional to square-root of the electron mass (see Eq.~(\ref{eqn:fermi_electron})).
\label{fig:mass_ratio}}
\end{figure*}

To further confirm that the initial energy gain of electrons is controlled by Fermi reflection on entry into reconnection exhausts, we explore the mass dependence of early time electron heating  with simulations employing varying electron-to-proton masses but with fixed upstream electron temperatures. In this situation, Eq.~(\ref{eqn:fermi_electron}) predicts that the initial electron energy gain should scale as $(m_e/m_i)^{1/2}$. Shown in Figure \ref{fig:mass_ratio} are cuts along the center of the current sheet of the electron parallel temperature at $t = 2.2\,\tau_A$ from three simulations, differing only by their mass ratio: $m_i/m_e = 25$, 100, and 400. The ion mass $m_i$ is fixed in all cases, so a higher mass ratio corresponds to a lower electron mass. At $t = 2.2\,\tau_A$ the electrons have interacted with a single reconnection outflow (see discussion related to Figure \ref{fig:early_time}). The cuts reveal that the simulation with $m_i/m_e = 25$ (heavier electrons) produces the strongest electron energy gain, while the case with $m_i/m_e = 400$ (i.e., lighter electrons) produces the least. To quantify this result, we have calculated the average parallel electron temperature increment within the outmost magnetic separatrix at $t=2.2\tau_A$ for each of the three simulations. The average temperature increments are 0.065, 0.045, and 0.038 for $m_i/m_e = 25$, $100$ and $400$, respectively. Thus, this early time temperature increment decreases with electron mass but not as much as predicted by the $(m_e/m_i)^{1/2}$ scaling of Eq.~(\ref{eqn:fermi_electron}). This is likely because, even at this early time, the high velocity electrons undergo many bounces in small magnetic flux ropes. This is especially true for the lower-electron-mass cases where the initial thermal speed greatly exceeds the Alfv\'en speed. In any case, the increased early heating of protons compared with electrons is a consequence of greater energy gain during Fermi reflection in developing reconnection exhausts.

Thus, the simulations suggest that protons gain substantially more energy than electrons during the early phase of reconnection and that their energy gain never overcomes that of the protons. That is, since the dominant mechanism for energy gain is Fermi reflection during flux rope merger, in which the rate of energy gain is proportional to energy \citep{Drake06a,Drake13}, once the proton energy exceeds that of electrons at early time the protons continue to gain energy faster and thus, reach higher energies over the full range of energies probed in the simulations. 

To further test this physical picture, we conducted a series of simulations varying the initial electron-to-proton temperature ratio while keeping the initial proton temperature constant. The idea is to boost the initial electron temperature to see if that would enable them to gain energy at a rate comparable to or greater than that of the protons. Specifically, the initial electron temperature was set to \( T_{e(\mathrm{initial})} = 0.5T_{i(\mathrm{initial})} \), \( T_{e(\mathrm{initial})} = T_{i(\mathrm{initial})} \), \( T_{e(\mathrm{initial})} = 2T_{i(\mathrm{initial})} \), and \( T_{e(\mathrm{initial})} = 5T_{i(\mathrm{initial})} \). At late time, typically after a single, large magnetic island dominates the current sheet, we computed the energy spectra of electrons and protons by aggregating particle counts across the entire simulation domain.

\begin{figure*}[ht!]
\centering
\includegraphics[width=\columnwidth]{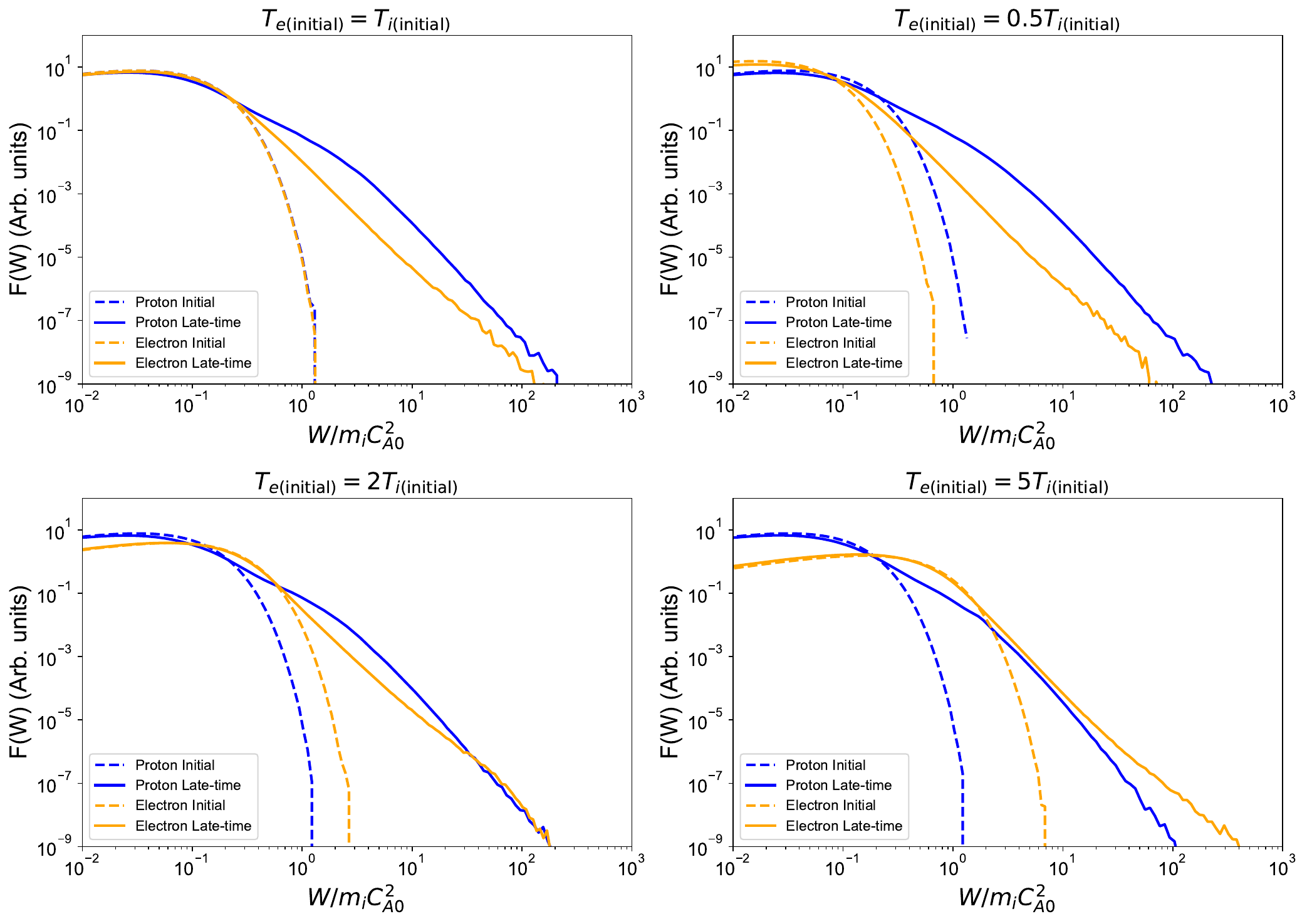}
\caption{
 Energy spectra of electrons and protons from four simulations with different initial electron-to-proton temperature ratios. The initial proton temperature is held constant, while the initial electron temperature is varied as $T_{e(\mathrm{initial})} = 0.5T_{i(\mathrm{initial})}$, $T_{e(\mathrm{initial})} = T_{i(\mathrm{initial})}$, $T_{e(\mathrm{initial})} = 2T_{i(\mathrm{initial})}$, and $T_{e(\mathrm{initial})} = 5T_{i(\mathrm{initial})}$, as indicated in each panel. The energy spectra are computed at late time, after the formation of a dominant magnetic island in the current sheet, by summing particles across the entire simulation domain. Dashed lines represent initial spectra, while solid lines indicate late-time spectra for protons (blue) and electrons (orange). At late time the spectral indices with the various initial electron temperatures differ only slightly from the equal temperature simulation. The powerlaw slopes of the proton spectra in these settings range from −4.0 to −3.6, whereas those of the electron spectra range from −3.2 to −3.0
\label{fig:spec}}
\end{figure*}

The resulting spectra, shown in Fig.~\ref{fig:spec}, reveal the influence of the initial temperature ratio on the relative energization of electrons and protons. In all cases, protons exhibit a extended high-energy, powerlaw tail at late time, consistent with strong energization via Fermi reflection. As the initial electron temperature increases, the late-time electron spectrum becomes progressively broader and more extended, indicating enhanced energization. In summary, particles with higher initial temperatures tend to reach higher energy regions in the final spectra. Notably, as the initial electron temperature increases, the proton spectra exhibit a slight decrease in overall energy gain, despite the proton temperature remaining fixed. This suggests a competition in energy partitioning: As electrons gain more energy as a result of their higher initial temperature, proton energization is somewhat suppressed.

To explore more fully the time evolution of particle energy gain, we show the space–time evolution of particle temperatures along the center of the current sheets in each of the simulations with differing initial electron temperature.  The goal is again to explore how the initial electron temperature impacts energy gain during reconnection. 

\begin{figure*}[ht!]
\centering
\includegraphics[width=\columnwidth]{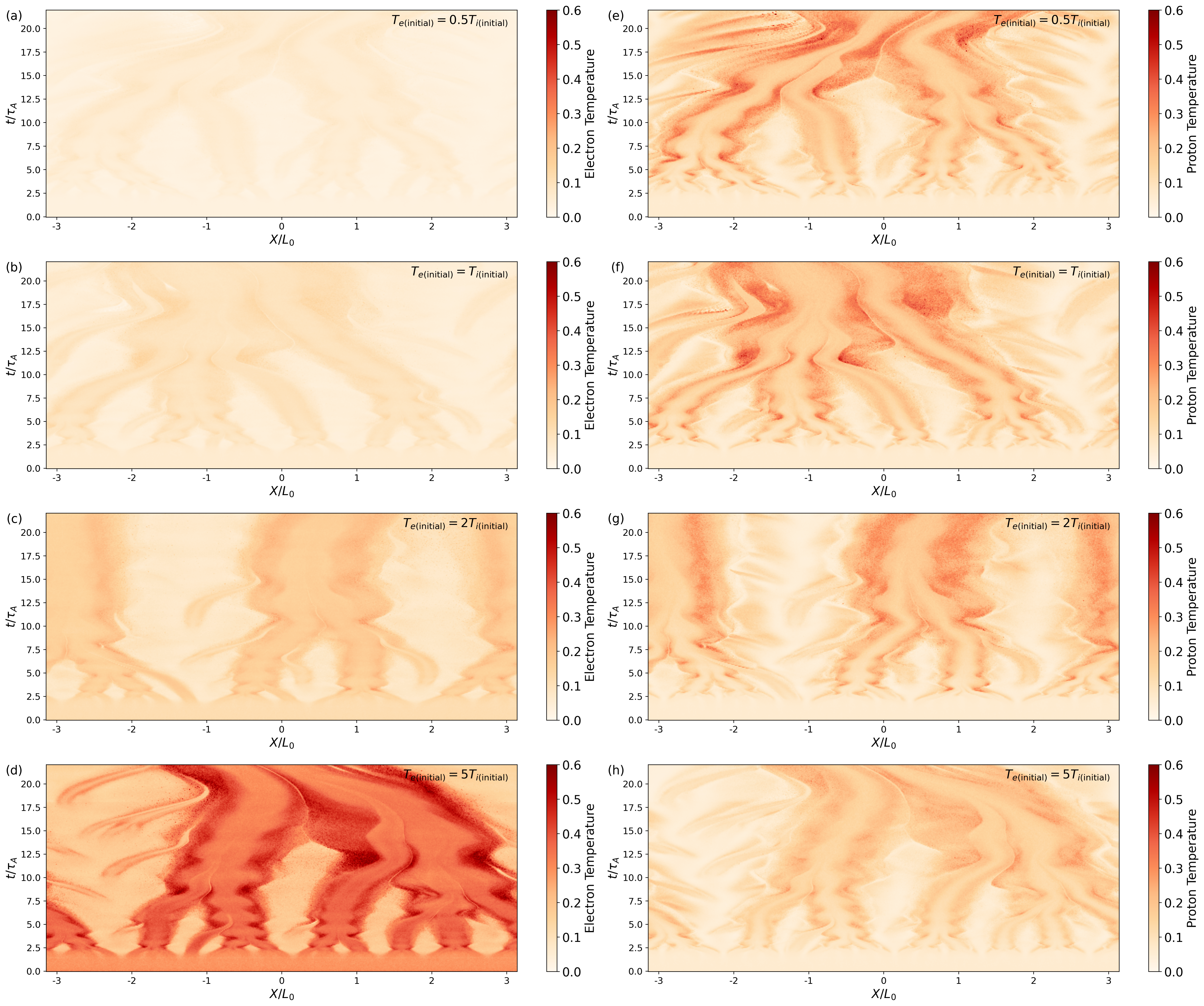}
\caption{Space–time diagrams of particle temperature along the center of the current sheet for different initial electron-to-proton temperature ratios. Panels (a--d) show the evolution of electron temperature, while panels (e–-h) display the corresponding proton temperature. The horizontal axis is the position \(X/L_0\) along the current sheet, and the vertical axis shows time  of \(t/\tau_A\) of the cut. The color scale indicates normalized temperature, with redder regions corresponding to higher temperatures. From top to bottom, each row corresponds to a different initial temperature ratio: \(T_{e(\mathrm{initial})}/T_{i(\mathrm{initial})} = \)1, 0.5, 2, and 5, respectively. The results show how both electron and proton heating profiles vary with the initial electron temperature. Higher initial electron temperatures yield stronger electron heating and reduced proton heating.
\label{fig:space_time}}
\end{figure*}

The space–time diagrams in Fig.~\ref{fig:space_time} clearly demonstrate how varying the initial electron-to-proton temperature ratio influences the heating of both species. As the initial electron temperature increases from $T_{e(\mathrm{initial})} = 0.5T_{i(\mathrm{initial})}$ to $T_{e(\mathrm{initial})} = 5T_{i(\mathrm{initial})}$, electron heating strengthens. A higher initial electron temperature enables the electrons to reach a higher temperature by simulation's end.  In particular, the case with $T_{e(\mathrm{initial})} = 5T_{i(\mathrm{initial})}$ exhibits widespread and persistent high-temperature regions even from very early times. The regions of nearly uniform temperature are at the cores of magnetic flux ropes, while the highest temperatures occur at the edges of the flux ropes. This is because the cores of flux ropes contain plasma that was heated early in time when reconnection outflows were weaker. Stronger outflows formed later in time and heated the plasma at the edges of flux ropes. Thus, these simulations establish that a higher initial electron thermal energy leads to more efficient and sustained electron heating and energization.

Interestingly, despite the proton initial temperature being fixed in all cases, the proton temperature maps (panels e-h) reveal systematic changes in response to the varying electron temperatures. At lower $T_{e(\mathrm{initial})}$, proton heating is relatively strong. However, as the initial electron temperature increases, the corresponding proton energy gain is smaller. This behavior suggests a redistribution of energy between the two species: enhanced electron energization suppresses proton heating and energization, even though the initial proton conditions remain unchanged. These observations underscore the coupled nature of energy conversion in collisionless magnetic reconnection and reinforce conclusions based on the particle spectra, in which increasing initial electron thermal energy boosted (suppressed) the production and maximum energy of electrons (protons).

\section{Conclusion} \label{sec:conclusion}

{\it In situ} observations of electron and proton heating during reconnection in near-Earth space have revealed that heating of both species scales with the available magnetic energy per particle $m_iC_A^2$, with protons gaining significantly more energy than electrons ($0.17\,m_iC_A^2$ for protons and $0.02\,m_iC_A^2$ for electrons \citep{Phan13a,Phan14,Oieroset23,Oieroset24}, although in the magnetotail observations there was some reduction of proton heating below this scaling for large values of $C_A$ \citep{Oieroset24}). This result was interpreted as arising from the electric potential that develops between the center of the reconnecting current sheet and the upstream, which can be substantial once the low density lobe plasma participates in the reconnection. PIC simulations of electron and proton heating during reconnection reproduce this basic trend \citep{Shay14,Haggerty15}. However, an understanding of why proton heating is much stronger than electron heating remains elusive. More recently, measurements have been extended to much higher energies, which are characterized by power-law distributions. There is some evidence that the power-law tails of protons contain more energy than those of electrons \citep{Ergun20b} and the first comprehensive study establishing this has recently been completed \citep{Rajhans25}. Further, in flare observations, where {\it in situ} measurements are not possible, the measurement of protons below an MeV are not available, so estimates of the relative energy content of the non-thermal components of the two species are uncertain \citep{Emslie12}. Recent simulations of electron and proton energy gain during reconnection, however, have revealed that non-thermal protons gain more energy than non-thermal electrons over several decades of energy \citep{Yin24b}. In the absence of observational support for this result during flare energy release, it is essential to establish the underlying physical mechanism responsible.  

In this paper, we have investigated proton and electron heating and energization during magnetic reconnection using the \textit{kglobal} simulation model. This is presently the only model that is able to produce the extended powerlaw energy spectra seen in observations. The simulations reveal that the protons experience a strong jump in energy early in time as they first enter into reconnection exhausts (Figures \ref{fig:evo}  and \ref{fig:space_time}). This energy jump greatly exceeds that of electrons. Since energy gain as time progresses is dominated by Fermi reflection in merging flux ropes, where energy gain is proportional to particle energy \citep{Drake06,Oka10,Drake13}, the initial strong heating of protons facilitates their increased energy gain compared with electrons for the duration of the reconnection dynamics. 

The initial energy gain of protons and electrons, which controls the subsequent dynamics, is controlled by the first entry of particles into the developing reconnection exhausts. As is well known, the protons counterstream at the Alfv\'en speed on their first entry into a reconnection exhaust and gain an energy that scales as $m_iC_A^2$. The simulations confirm this result (see Figures \ref{fig:space_time} and \ref{fig:early_time}). In contrast, because of their smaller mass electrons gain an energy that scales as $(\beta_{e0}m_e/m_i)^{1/2}m_iC_A^2$. Simulations with varying electron mass ($m_i/m_e=25$, 100, 400) have been completed that confirm that the initial electron temperature gain on exhaust entry decreases with electron mass. However, the precise scaling is ambiguous because of the turbulent nature of the early reconnection dynamics.

To further confirm that it is the initial injection energy that controls the subsequent relative energization of the two species, we carried out a series of simulations in which we varied the initial upstream electron temperature. In a simulation with $T_{e0}/T_{i0}=0.5$ the proton heating and energization further increased compared with the case with equal upstream temperatures (see Figures \ref{fig:spec} and \ref{fig:space_time}). In a series of simulations in which $T_{e0}/T_{i0}$ was progressively increased, electron energization became progressively stronger. Electron energy gain exceeded that of protons across the full energy spectrum for $T_{e0}/T_{i0}=5.0$. 

Beyond the overall energy spectrum of electrons and protons, we have focused on obtaining an explicit value for the electron and proton increment during reconnection. We emphasize that since the energy normalization of the \textit{kglobal} model is $m_iC_A^2$ the scaling of heating of both species scales with this parameter, as documented in observations. Since the value of the electron mass in the simulations is artificial, the sensitivity of the result to $m_i/m_e$ was checked with simulations with mass ratios $m_i/m_e=25$, $100$ and $400$. The electron and proton temperature increments for simulations with these three mass-ratios are, respectively: $\Delta T_e= 0.070$, $0.079$ and $0.071$; and $\Delta T_i=0.313$, $0.306$ and $0.314$. Thus, the late time relative temperature increments of the two species are insensitive to the mass-ratio, a result that is consistent with previous PIC simulations \citep{Shay14}. This result would seem to be at odds with the conclusions of \cite{Oka25} who concluded that the ratio of proton to electron heating scales as $(m_p/m_e)^{1/4}$ based on the analysis of {\it in situ} observations. However, the mass ratio in the observations was clearly 1836 and was not varied. Thus, the observation of a temperature ratio of around 6.5 is an empirical result, rather than the result of a scaling with $m_p/m_e$. Finally, we emphasize that the increased energy increment of protons compared to electrons persists at high energy in the non-thermal spectra of the two species.  

Extensive {\it in situ} satellite measurements of particle temperature increments during reconnection scale as $\Delta T_e = 0.02m_iC_A^2$ and $\Delta T_i = 0.13m_iC_A^2$. Thus, there are differences between the present simulation results and the observational data. The proton temperature increments in the simulations are nearly a factor of two larger than in the data, although, as discussed previously, the simulations support the scaling with $m_iC_A^2$. The ratio of the proton and electron temperature increments are about a factor of four in the simulations, compared with around six in the data. This may be a consequence of the higher upstream temperature of protons versus electrons (by around a factor of five) in the Earth's magnetosphere, where most of the observations were taken. The reasons for the stronger heating in the simulations are not fully understood. However, the present simulations focus on multi x-line reconnection which lead to extended power-law distributions of both species. It is possible that multi x-line reconnection leads to larger temperature increases than reconnection at a single x-line, observations of which likely dominate the spacecraft data. The idea that multi x-line reconnection could lead to higher temperature increments is consistent with the cuts of the early time data shown in Fig.~\ref{fig:early_time}(c). The peak parallel temperature increments of the protons due to a single entry into an exhaust are around 0.4$m_iC_A^2$. Since the perpendicular temperatures remain nearly unchanged (or even slightly reduced), the total temperature increment of protons on single entry is about one third of this value or around 0.13$m_iC_A^2$, which is consistent with the observational data. 

A recent survey of energetic reconnection events in the Earth's magnetotail provides important observational constraints on eletron and proton heating and acceleration \citep{Rajhans25}. Consistent with the present simulations, these observations revealed that protons consistently gain more energy than electrons across the full range of particle energies (in the thermal and non-thermal components of both species). However, there are differences between the observations and modeling results. The observations revealed that electrons typically have harder spectra even though the non-thermal electrons carried less energy than the non-thermal protons. In the present simulations the spectral indices of the non-thermal particles are comparable (see Fig.~\ref{fig:spec}) even when the initial ratio of electron and proton temperatures varies. The electron spectra at high energy are slightly harder than those of the protons. A clear difference between the simulations and the observations is the measured (in observations) inverse correlation between the fraction of non-thermal particles and the measured temperature of those particles. In the present model, the non-thermal fraction of both species depends strongly on the strength of the guide field and not the value of $m_iC_A^2$ (which controls the temperature) \citep{Arnold21,Yin24b}. On the other hand, the present model does not include some potentially important physics. In the simulations there is no mechanism for energetic particles to escape from the energy release region, a process that could limit the energy content of the non-thermal particles. Finally, the demagnetization of protons at high energy, which is also not included in the {\it kglobal} model, might limit the energy gain of the most energetic non-thermal protons. Further exploration of these topics in the models is warranted.  

The simulations presented in this manuscript are based on the assumption that classical collisions can be neglected. In the case of the solar wind and the Earth space environment such an assumption is reasonable because of the very low plasma densities and associated weak classical collisions. In the case of solar flares, however, the neglect of classical collisions is not as obvious. A key parameter is the strength of the reconnection electric field compared with the Dreicer runaway electric field $E_D$, which is the electric field needed to accelerate electrons to their thermal speed during a classical collision time \citep{Dreicer60}, 
\begin{equation}
    eE_D=m_e\nu_ev_{the}
    \label{eqn:E_D},
\end{equation}
with $v_{the}$ the electron thermal speed and $\nu_e$ the electron collision rate, 
\begin{equation}
    \nu_e=2.9\times 10^{-6}n_e\ln\Lambda/T_e^{3/2}
    \label{eqn:nu_e},
\end{equation}
where $\ln\Lambda$ is the Coulomb logarithm, the density $n_e$ is in units of $cm^{-3}$ and the electron temperature is in $eV$. For typical coronal ambient parameters ($n_e\sim 10^{10}/cm^3$ and $T_e\sim 100eV$), $\nu_e\sim 440/s$ and $E_D\sim 1.05\times 10^{-2}V/m$. In the collisionless limit the reconnection electric field $E_r$ is around $0.1C_AB$, where $C_A$ is the upstream Alfv\'en speed \citep{Shay99,Shay07,Torbert18,Burch20}. For a magnetic field of 50$G$ with the same density, $C_A\sim 1.1\times 10^6m/s$ and $E_r\sim 550V/m$, which is far above the Dreicer value. Thus, reconnection in solar flares is deeply in the collisionless regime. 

On the other hand, the dominant driver of the energization of the highest energy electrons and ions in flares is the merger of flux ropes. The merger of two equal-sized flux ropes causes the energy of all particles within the flux ropes to approximately double (the field line shortens by a factor of $\sqrt{2}$ and, as a result of the second adiabatic invariant, the particle parallel velocity increases by the same factor) \citep{Drake13}. Classical collisional energy losses can be neglected if the merging time $\tau_r$ is less than $\nu_e^{-1}$. The merging time of two flux ropes of radii $R$ is therefore given by $10R/C_A$ and the neglect of classical energy loss requires
\begin{equation}
    \nu_e<0.1C_A/R.
    \label{eqn:merger}
\end{equation}
Reconnection should onset with the smallest flux ropes that can magnetize protons or $R\sim5d_i$ with $d_i=c/\omega_{pi}=C_A/\Omega_{ci}$ the ion inertial length scale \citep{Mandt94}. Thus, the neglect of collisional losses requires
\begin{equation}
    \nu_e<0.02\Omega_{ci}
\end{equation}
with $\Omega_{ci}$ the proton cyclotron frequency. 
This inequality is easily satisfied for the typical solar parameters previously discussed. Further, the inequality in Equation (\ref{eqn:merger}) will continue to be satisfied during subsequent island mergers. The  collision rate scales as $v^{-3}$. Since $v$ increases by the factor $\sqrt 2$ during each merger, after $N$ mergers $\nu_e$ is reduced by $2^{-3N/2}$. The merging time $\tau_r$, however, increases with each merger: $\tau_r=10R/C_A\propto R/B\sim R^2\sim 2^N$ since the magnetic flux $BR$ is invariant during island merger \citep{Fermo10}. Thus, $\nu_e\tau_r \propto 2^{-N/2}$. If Equation (\ref{eqn:merger}) is satisfied, $\nu_e\tau_r<1$ for all subsequent mergers as particles gain energy. 

Also of interest in solar flares is the rate at which particles gain energy and whether it is fast enough to be consistent with the rise time of hard x-rays. The shortest time scales are likely to be associated with island merger. The merger time depends, of course, on its radius as well as the local upstream Alfv\'en speed. Since a range of island sizes are expected in a reconnecting current layer \citep{Fermo10}, there is no universal time associated with island merger.  For an island with scale size around 100$km$ with $C_A\sim 1000km/s$ as discussed above, the merger time and therefore the energy doubling time is around one second. Smaller islands can merge more quickly and are also able to drive strong particle energy gain since even the Larmor radii of protons in the MeV range are only hundreds of meters and are therefore well magnetized.  


\begin{acknowledgments}
We acknowledge extensive discussions with Drs. Tai Phan, Marit {\O}ieroset, Mitsuo Oka and the MMS Science Team. Support was provided from NASA Grant Nos. 80NSSC20K1277, 80NSSC20K1813 and
80NSSC22K0352, and NSF Grant No. PHY2109083. The
simulations were carried out at the National Energy Research
Scientific Computing Center (NERSC). The data used to perform
the analysis and construct the figures for this paper are preserved at the NERSC High Performance Storage System and are available upon request.  
\end{acknowledgments}

\bibliography{sample7}{}
\bibliographystyle{aasjournalv7}

\end{CJK*}
\end{document}